\documentclass[aps,prl,twocolumn,showkeys,amsmath,amssymb,longbibliography,floatfix]{revtex4-1}
\usepackage[T1]{fontenc}
\usepackage{graphicx}
\usepackage{dcolumn}

\usepackage{amsmath,amssymb,amsfonts,bm,dsfont,units}
\usepackage{hyperref}
\usepackage{tabularx}
\hypersetup{colorlinks=true,citecolor=blue,linkcolor=blue,urlcolor=blue,pdfstartview=FitH}
\usepackage{appendix} 
\usepackage{braket}
\usepackage{float,latexsym}
\usepackage{multirow}
\usepackage{diagbox}
\usepackage[normalem]{ulem} 
\usepackage[caption=false]{subfig} 
\usepackage{comment}
\usepackage{tikz}
\usepackage{xcolor}

\usepackage{array}
\usepackage{graphicx} 
\usepackage{hhline,booktabs}
\usepackage{makecell}
\usepackage{wasysym} 
\usepackage{array}
\newcolumntype{Y}{>{\centering\arraybackslash}X}

\begin{document}

\title{The Anyon Zeno Effect}

 \author{David F. Mross}
\affiliation{Department of Condensed Matter Physics, Weizmann Institute of Science, Rehovot 7610001, Israel}
 
\begin{abstract} 
Two anyons encircling each other acquire a quantized braiding phase that is independent of their spatial separation. We show that detecting this phase in a fractional quantum Hall interference experiment results in a quantum Zeno effect: a localized anyon is trapped by constant observation from a stream of anyons supplied by the measurement current. Interferometers with an embedded antidot are ideal for accessing the Zeno regime, where the bare tunneling rate of localized anyons is much lower than the measurement rate. The Zeno-suppressed tunneling rate of the trapped anyon depends on the braiding phase and the transmission of the quantum point contacts. Our primary prediction is that the autocorrelation time of the conductance through the interferometer increases with the bias current. This effect may be used to experimentally control the anyon dynamics, in particular to increase the lifetime of localized anyons.
\end{abstract}

\date{\today}
\maketitle

\textit{Introduction}.~Point-like particles in two dimensions can satisfy anyonic statistics \cite{Leinaas1977,Wilczek1982,stern2008}. Their wavefunction does not return to its original form after one particle has encircled the other, as it does for bosons or fermions, but acquires a braiding phase $e^{2 i \theta}$. A direct consequence of anyonic braiding is that these particles can function as each other's observers over arbitrary distances, purely through their statistics. The phase one anyon (`Alice') acquires upon encircling a second anyon (`Bob') could be detected via interference, and thus constitutes a measurement of Bob's presence. 

If Bob is originally in a quantum superposition between states on either side of Alice's trajectory, his wavefunction collapses as a result of her observation. Repeated measurements made by Alice can, therefore, disrupt Bob's tunneling process and keep him trapped over long time scales. This general phenomenon is known as the quantum Zeno effect \cite{MisraSudarshan1977,Itano1990}; see Ref.~\cite{greenfield2025} for a recent review. Bob's lifetime increases linearly with the measurement rate and diverges in the limit of continuous projective measurement. His immobility due to Alice's observation is a counterpart of `which path' detection \cite{aleiner1997,Buks1998}, where 
Alice's trajectory is measured, resulting in dephasing and a loss of interference. 

Ideal platforms for observing such an anyon Zeno effect are fractional quantum Hall interferometers, where anyonic statistics have been demonstrated \cite{nakamura2020direct,nakamura2023fabry,kim2024aharonov, werkmeister2024anyon, samuelson2024anyonic,Ghosh_OMZI_2024, Ghosh_Coherent_Bunching_2024,kim2025aharonov}. The two architectures used in those experiments are the Fabry-P\'erot Interferometer (FPI) \cite{Chamon1997} and the optical Mach-Zehnder Interferometer (OMZI) \cite{giovannetti2008,deviatov,batra2023}.
In both interferometers, the fractionally charged anyon Alice emitted from the source $S$ can tunnel through the quantum point contacts (QPCs) and flow along gapless edge states on either side of the central region. The two paths differ by a closed loop around the center, i.e., by one winding around the localized anyon Bob. Alice's probability to arrive at either drain depends on the braiding phase acquired in such a loop, so that a measurement of the conductance between the source $S$ and either drain reveals Bob's presence 
\nocite{zhang2009,ofek2010,halperin20211,Levkivskyi2012}
\footnote{For the FPI, interference can be overshadowed by Coulomb charging effects \cite{zhang2009,ofek2010,halperin20211,Levkivskyi2012}, but we will assume the Aharonov-Bohm regime realized in recent experiments \cite{nakamura2020direct,nakamura2023fabry,kim2024aharonov, werkmeister2024anyon, samuelson2024anyonic,kim2025aharonov}. This complication is absent in OMZIs, where the equal chirality of the top and bottom edges ensures each particle passes through the interferometer.}.
\begin{figure}
 \centering
 \includegraphics[width=\linewidth]{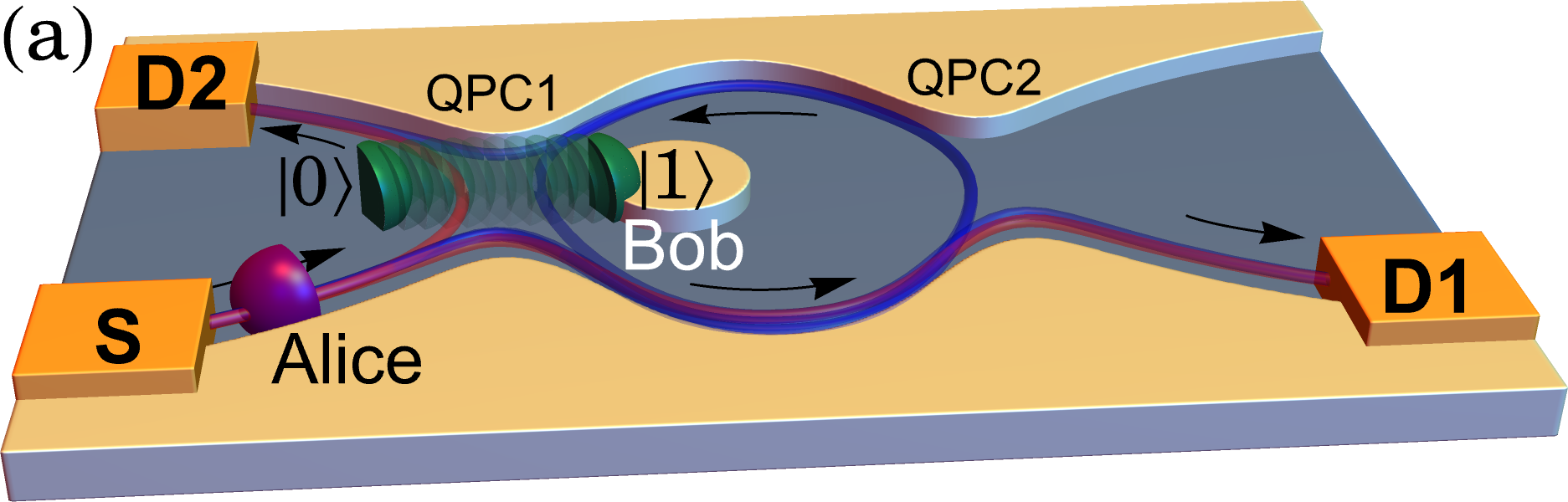}
\includegraphics[width=\linewidth]{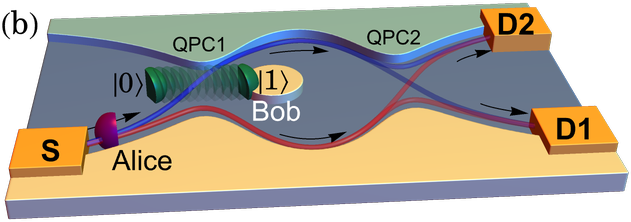}
 \caption{Fabry-P\'erot Interferometer (a) and Optical Mach-Zehnder Interferometer (b) with embedded quantum antidots. The localized anyon `Bob' (green) can tunnel between states $|1\rangle$ in the center and $|0\rangle$ outside the loop. His tunneling is disrupted by the interfering anyon `Alice' (purple), propagating from the source $S$ to the drains $D1$, $D2$, along edge states with the indicated chirality.
 }
 \label{fig.phase_quart}
\end{figure}

We propose that the anyon Zeno effect can be observed by sourcing currents $I$ in the nanoampere range into FPIs or OMZIs with embedded antidots. Such devices are illustrated in Fig.~\ref{fig.phase_quart} and were recently realized in GaAs \cite{Ghosh_OMZI_2024, Ghosh_Coherent_Bunching_2024,kim2025aharonov} and bilayer graphene \cite{KimAD,YuvalAD}. There, Bob's escape time can be obtained by recording the conductance $G(t)$ over time. A Poissonian process of switching between Bob's states leads to random telegraph noise characterized by exponentially decaying autocorrelations
 \begin{align}
 C(\Delta t)\equiv \overline{ G(t) G(t+\Delta t)} - \overline{G(t)} ^2 \sim e^{-2 |\Delta t| \Gamma_\mathrm{anyon}},
 \label{eqn.autocorr}
\end{align}
where the overline denotes a time average.  We predict that the escape rate $\Gamma_\mathrm{anyon}$ is inversely proportional to the bias current, which sets the rate of Alice's measurements. The anyon Zeno effect can also occur without an antidot, and the slow quasiparticle dynamics reported in Refs.~\cite{werkmeister2024anyon, samuelson2024anyonic} may be a manifestation of this phenomenon.

\textit{Effective description as a two-level system}.~The antidot inside an FPI or an OMZI can localize the anyon Bob with a small amplitude to tunnel outside of the interference loop. His two possible states, $|1\rangle$ inside the loop and $|0\rangle$ outside the loop, form a two-level system, described by
\begin{align}
 H_\mathrm{Bob} = \hbar \Omega_\Delta \sigma_z+\hbar \Omega \sigma_x~,\label{eqn.bob} 
\end{align}
with Pauli operators $\sigma_{x,z}$. The first term encodes the energy difference between $|0\rangle$ and $|1\rangle$, and the second parameterizes the tunneling between them. The latter is exponentially small in the antidot-edge distance, and can be much smaller than the measurement rate $\gamma_M = I/e^*$, where $e^*$ is Alice's charge.

For repeated projective measurements, Bob's lifetime could be read off from the unitary time evolution governed by $H_\mathrm{Bob}$ between measurements. One finds the well-known Zeno and anti-Zeno transition rates:
$ \Gamma_\mathrm{Zeno} = 
 \Omega^2 \gamma_M^{-1}$ for $\gamma_M\gg \Omega_\Delta$ and $\Gamma_\mathrm{anti-Zeno} =
 \frac{\Omega^2}{\Omega_\Delta^2} \gamma_M $ for $ \gamma_M \ll \Omega_\Delta $ \cite{MisraSudarshan1977,LANE198491,SCHIEVE1989264,Itano1990,facchi2001,KofmanKurizki2000, FacchiPascazio2008}. However, the anyon Zeno effect differs from the simple model of a projectively measured two-level system in two significant ways. Firstly, the measurement is generally not projective. More significantly, Bob can undergo unitary evolution due to Alice's passage in addition to the measurement backaction.

\begin{figure}
 \centering
\includegraphics[height=2.55cm]{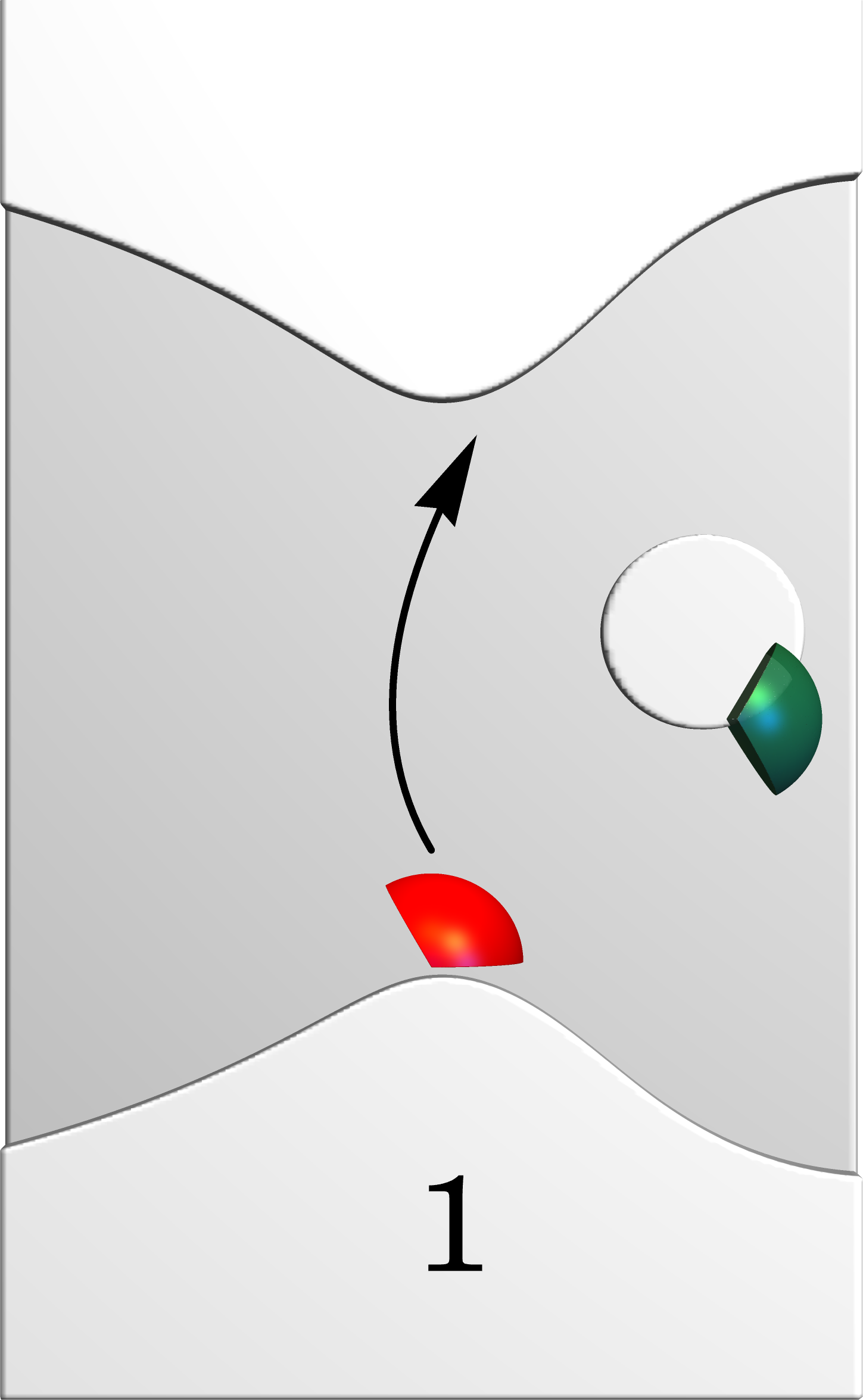}
 \includegraphics[height=2.55cm]{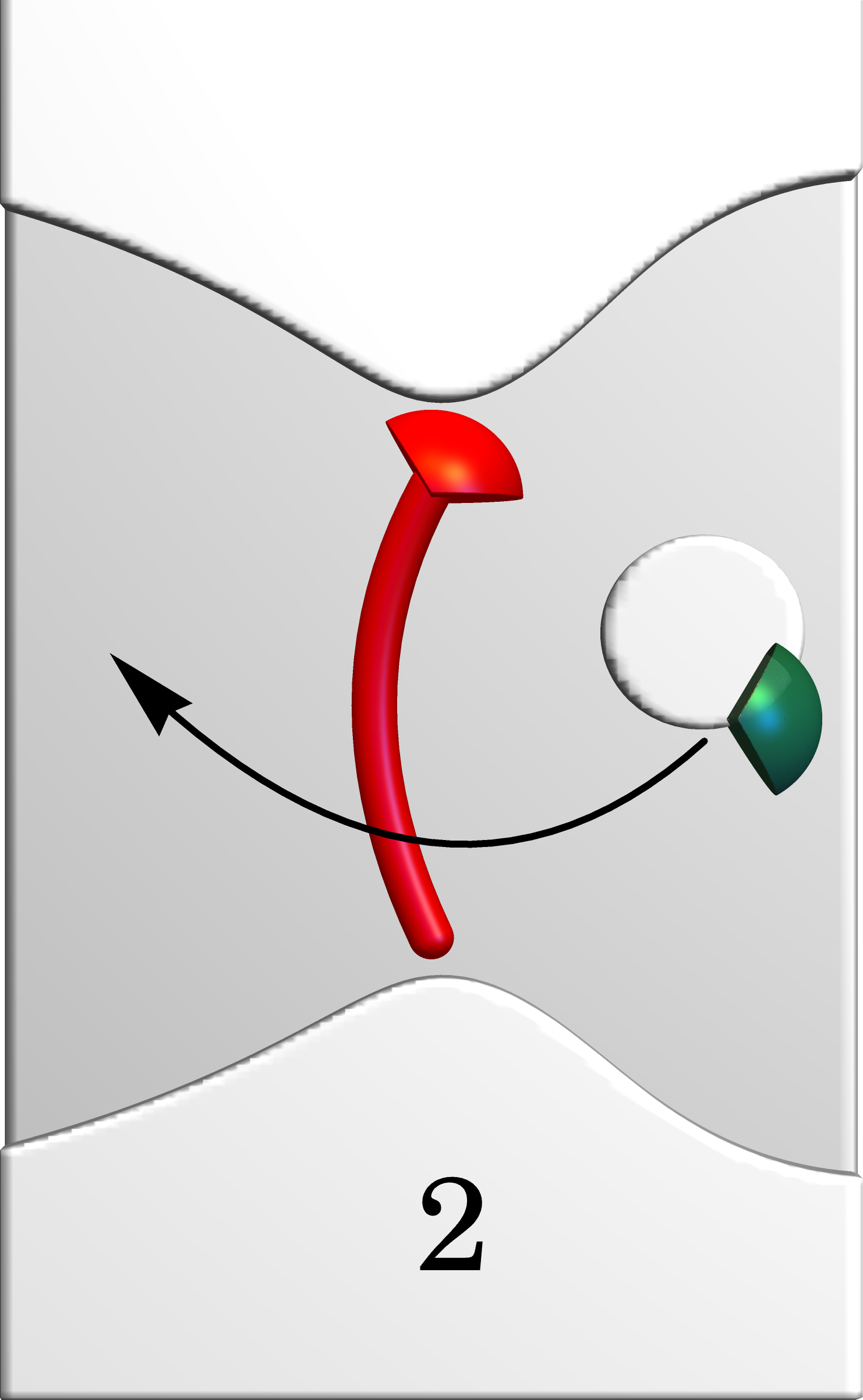}
 \includegraphics[height=2.55cm]{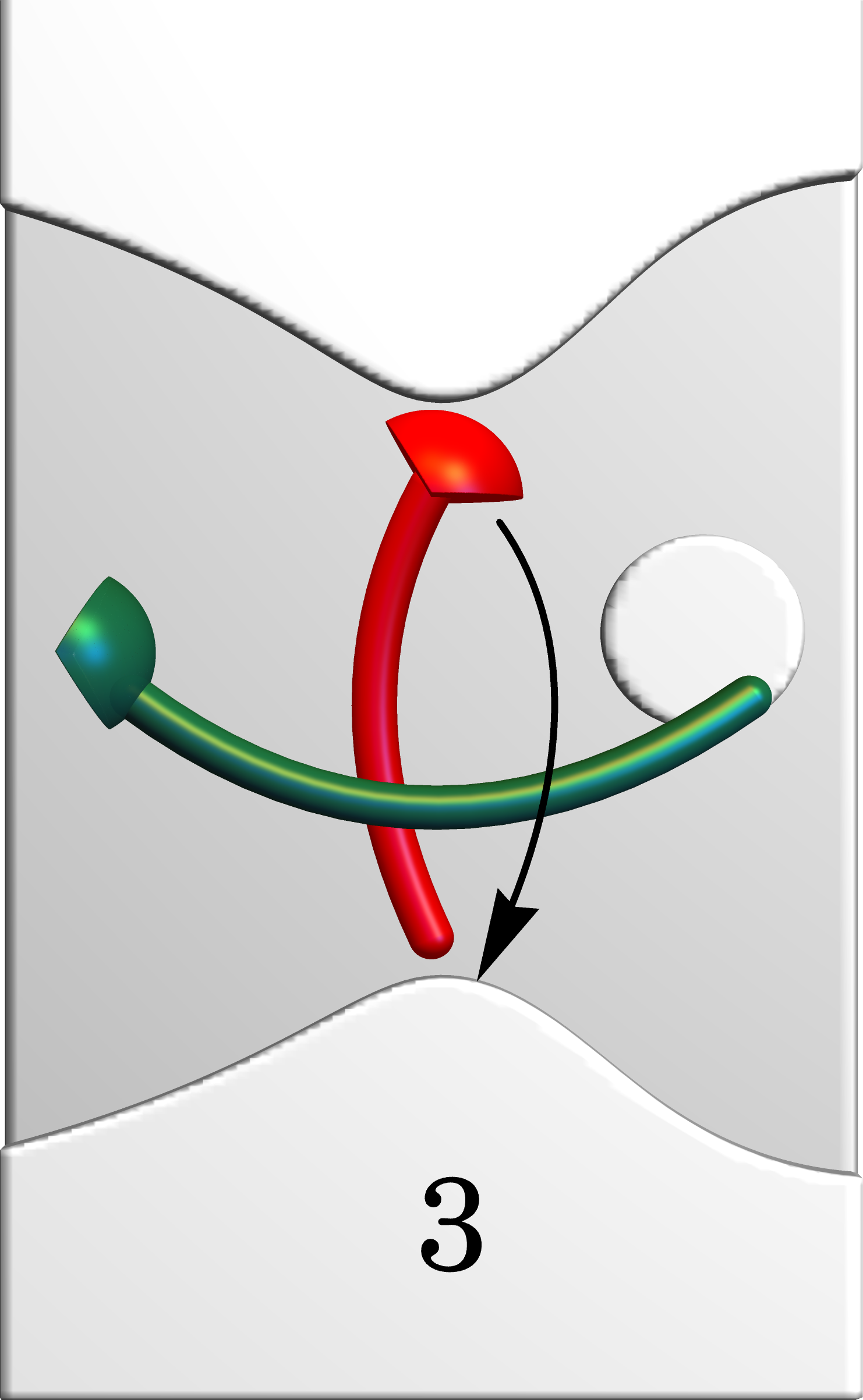}
 \includegraphics[height=2.55cm]{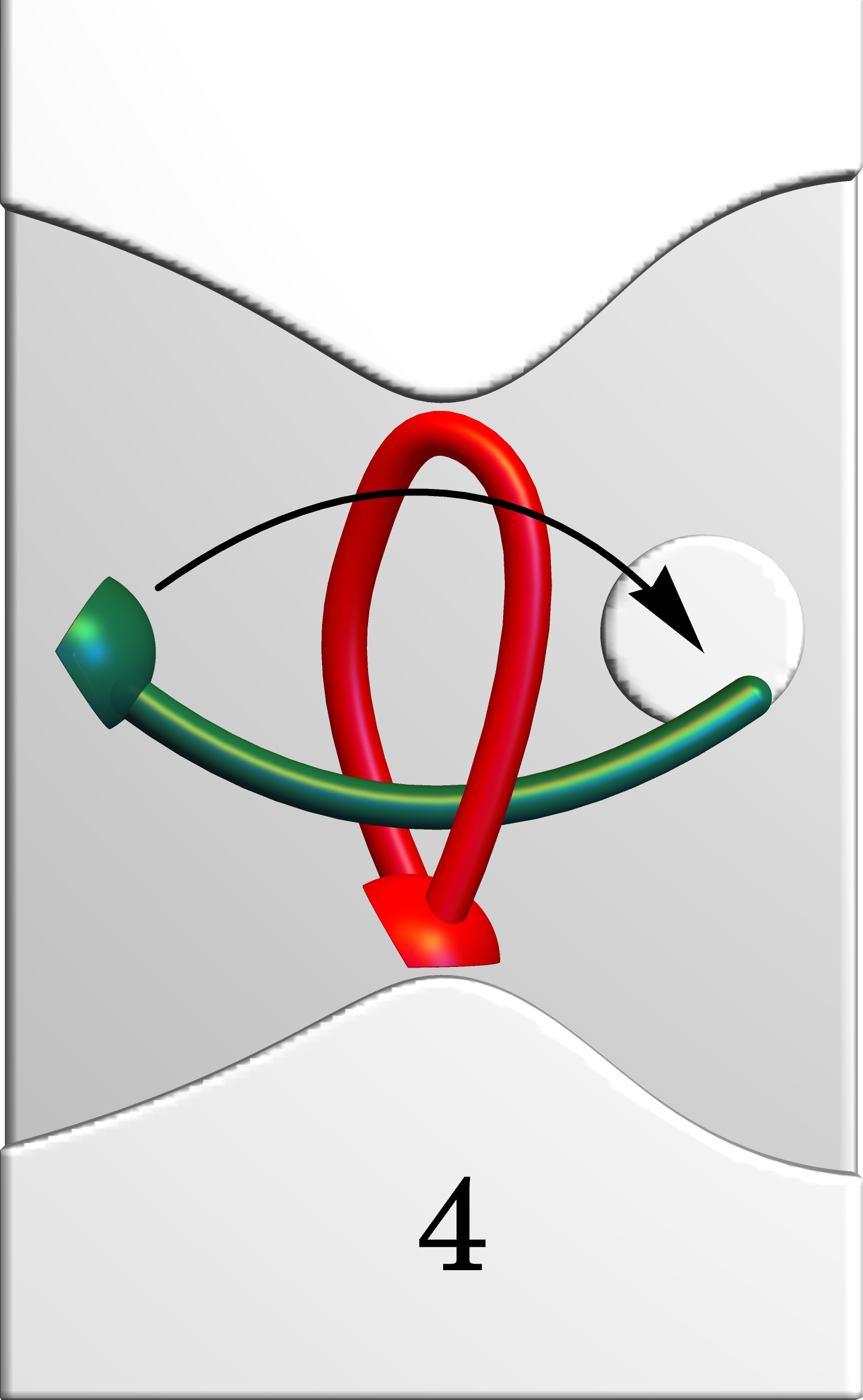}
\includegraphics[height=2.55cm]{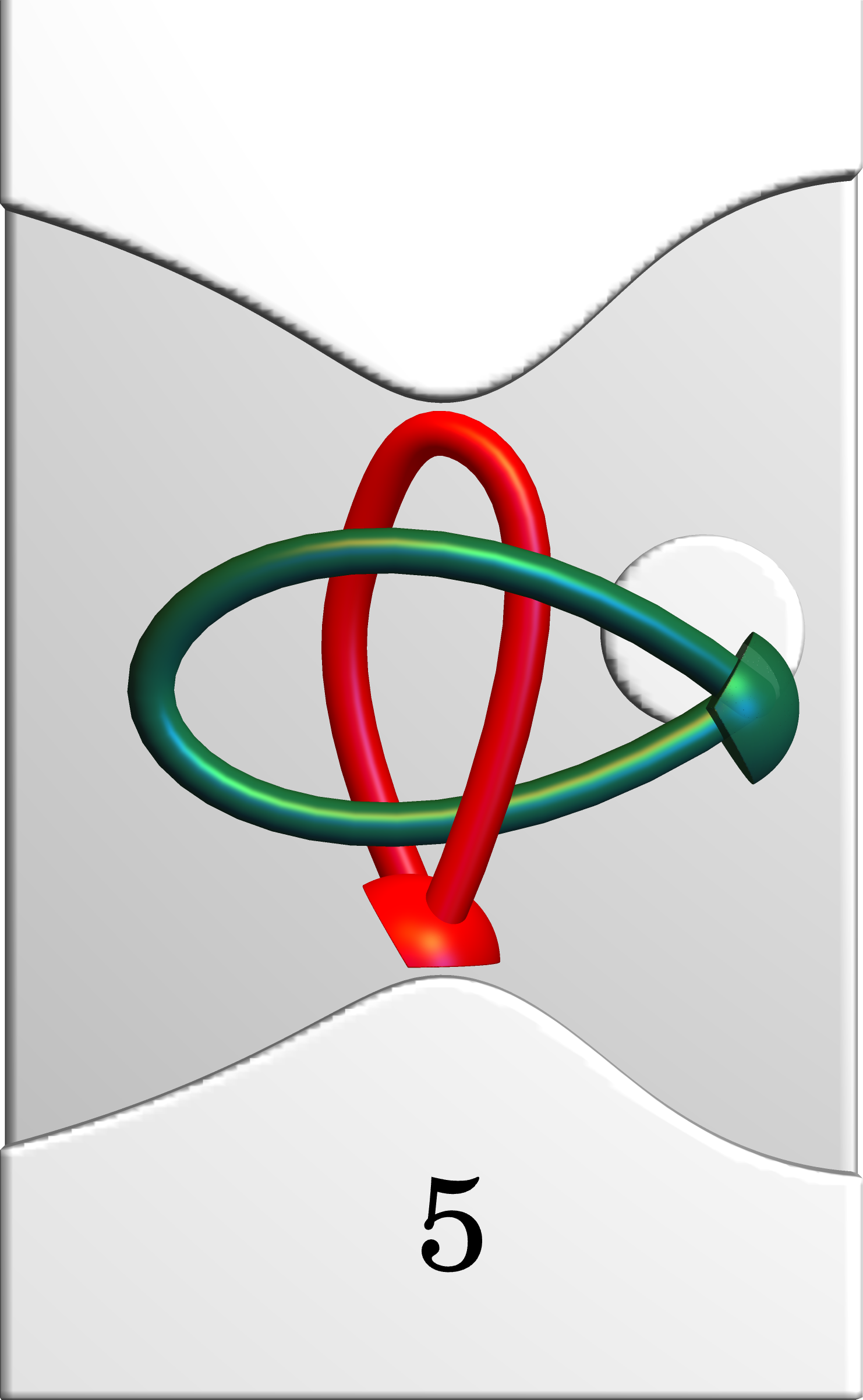}\\[2mm]
\includegraphics[height=3.0cm]{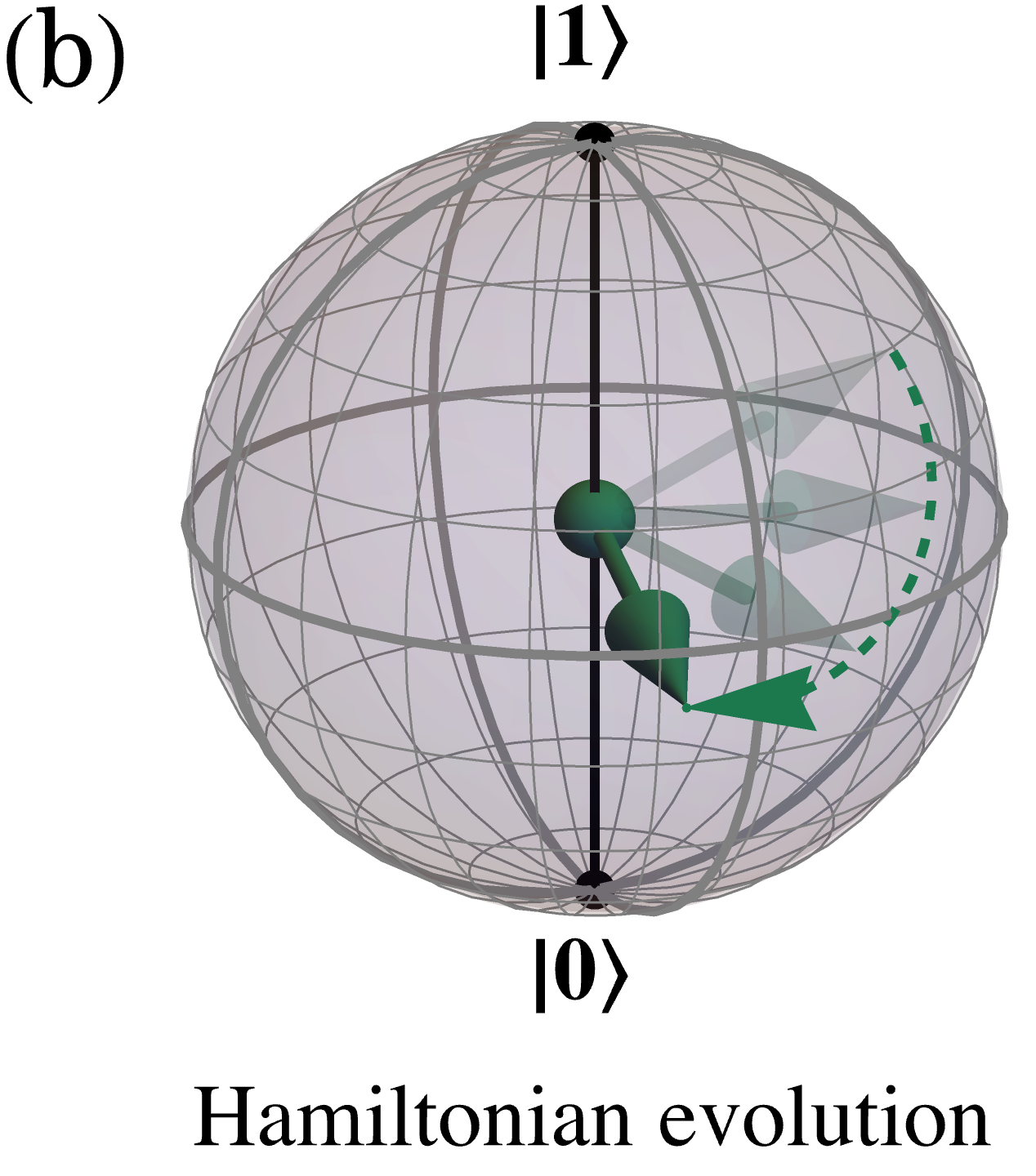}
\hspace{2mm}
\includegraphics[height=3.0cm]{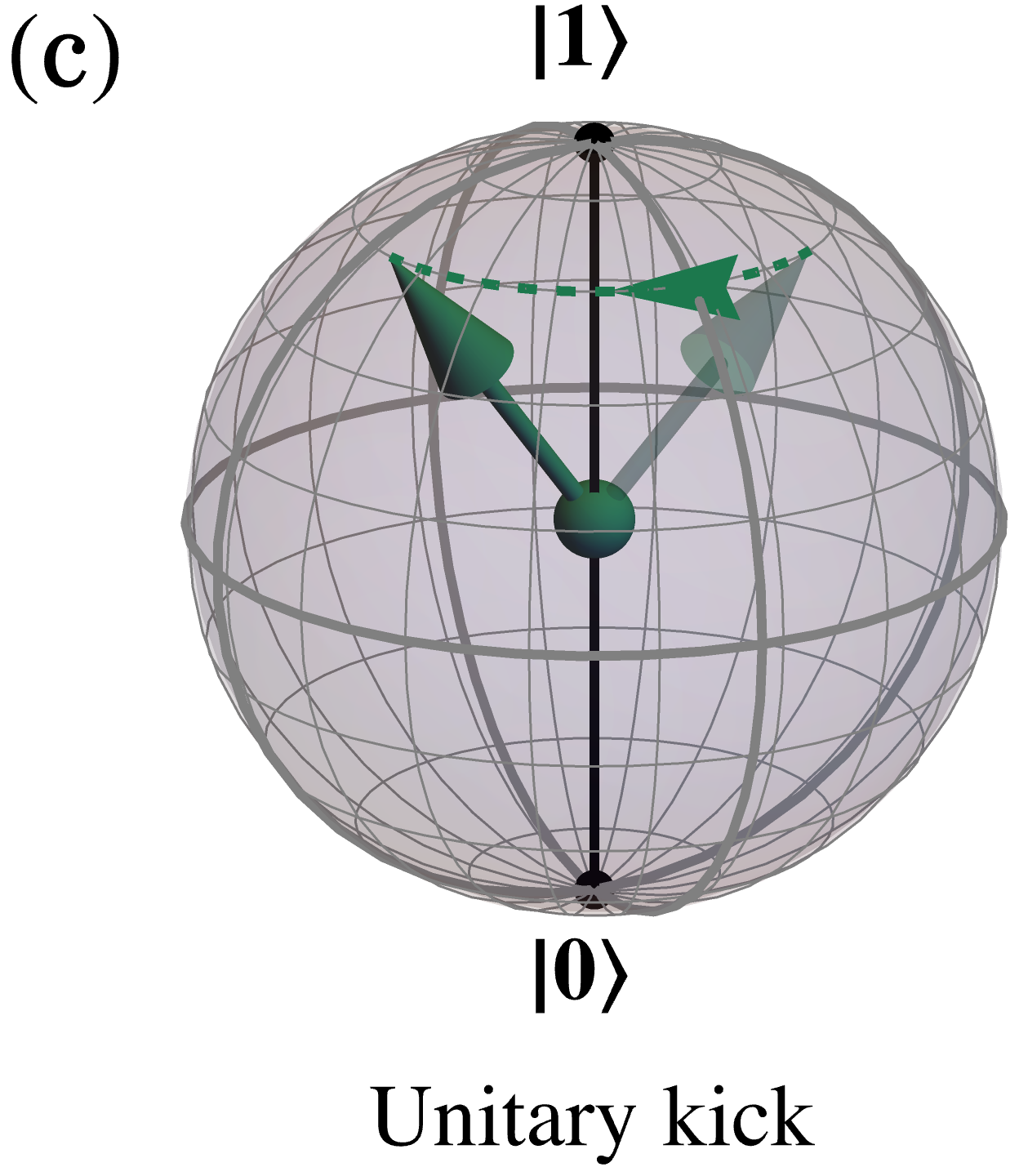}
\hspace{2mm}
\includegraphics[height=3.0cm]{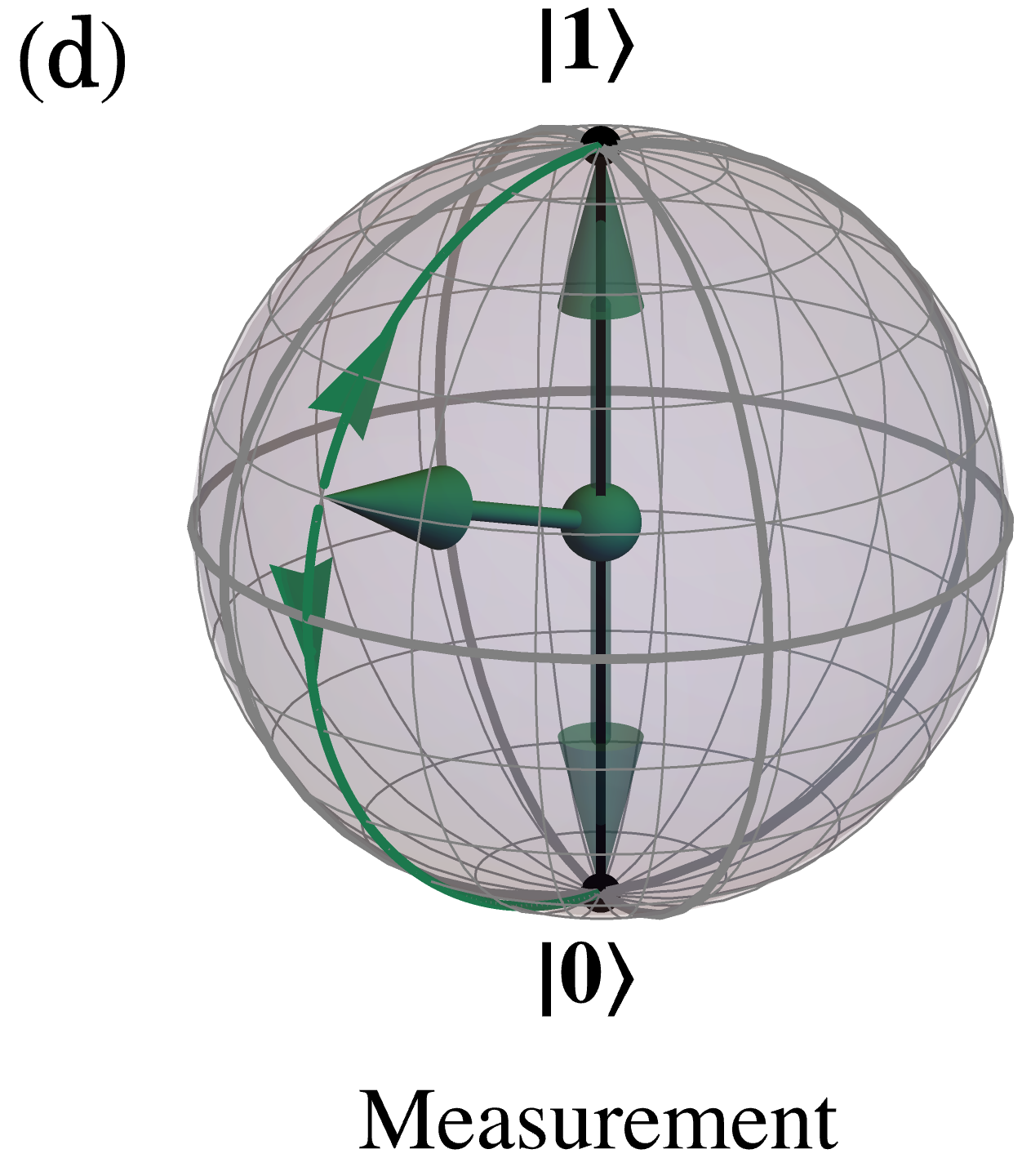}

 \caption{(a) A tunneling sequence of Alice and Bob  at QPC1 that can be viewed as time-domain braiding \cite{Han2016}. This process interlinks their worldlines and is associated with a phase $e^{2 i \theta}$. (b)-(d) Three mechanisms determine Bob's dynamics: Continuous Hamiltonian time evolution, unitary kicks due to Alice's passage and measurements that partially project Bob onto the states $|0\rangle$ or $|1\rangle$.}

 \label{fig.braiding}
\end{figure}

\textit{Anyon tunneling and braiding at a QPC}.~
We analyze Bob's dynamics under the assumption that Alice and Bob interact solely through their statistics. When both QPCs are at full transmission, Bob's time evolution is governed by Eq.~\eqref{eqn.bob} and can be represented by a Bloch vector rotating continuously around a fixed axis. As QPC1 is partially closed, Alice can tunnel to the opposite side, which we describe by the scattering matrix \begin{align}\hat S_1 =\begin{pmatrix}
 t_1 & \hat r_1 \\ -\hat r_1^\dagger & t_1
\end{pmatrix}\qquad 
\hat r_1 = r_1 e^{ i2 \theta \hat N}~,
\label{eqn.smatrix}
\end{align}
where $t_1$ is real, and the anyon number $\hat N\equiv (1+\sigma_z)/2$ is one (zero) when Bob is inside (outside) the interferometer. To understand the appearance of this operator, we consider the process illustrated in Fig.~\ref{fig.braiding}(a). Bob is initially inside the loop, and Alice tunnels from the lower to the upper edge. Next, Bob leaves the loop, and Alice returns to the lower edge. Finally, Bob returns to his original position. Through this process, Bob has performed a closed space-time braiding loop around Alice, i.e., $ \sigma_+ \hat r_1^{-1} \sigma_- \hat r_1 = \hat N e^{2 i \theta}$, yielding Eq.~\eqref{eqn.smatrix}. In contrast, Alice's transmission past the QPC along the bottom edge does not result in any braiding.

To analyze the implications of Eq.~\eqref{eqn.smatrix} for Bob's time evolution, we consider Alice starting at the source in state $|S\rangle$ and impinging on the QPC with Bob in state $|B\rangle$. In the initial state, $|i\rangle =|B\rangle \otimes |S\rangle $, there is no entanglement between the two. After Alice has passed the QPC, their joint state
\begin{align}
\begin{split}
 |f\rangle =\hat S_1 |i\rangle=&t_1 |D_1\rangle\otimes |B \rangle +
 r_1 |D_2\rangle \otimes |e^{i 2 \theta \hat N} B\rangle
\end{split}
\label{eqn.out}
\end{align}
is entangled due to the $\hat N$ operator in the second term. Bob's reduced density matrix $\rho =\mathrm{Tr}_{D_{1,2}}|f\rangle \langle f|$ describes a pure state only in three cases: (i) There is no partitioning, $t_1^2=0$ or $1$. (ii) The braiding phase is trivial, $\theta=0$ or $\pi$. (iii)  Bob is entirely inside or outside, such that $\hat N$ reduces to a number. In all other cases, there is a measurement backaction on Bob and $\mathrm{Tr} \rho^2 <1$.

Alice's probability of arriving at drain $D_i$ is given by $P_i=|\langle f |D_i\rangle|^2$. Using the state $|f\rangle$ of Eq.~\eqref{eqn.out}, one finds that this probability is independent of $|B\rangle$; the source-drain conductance cannot reveal Bob's presence while QPC2 is open. This observation does not contradict the statements we made about Alice's measurement of Bob in the previous paragraph. Their entanglement is encoded in the phase difference between the two terms in Eq.~\eqref{eqn.out}, which does not affect $P_i$.

\textit{Zeno effect at a QPC}.
When Alice propagates to $D_2$, Bob's Bloch vector rotates by $2\theta$ around the $\hat z$-axis. His Hamiltonian time evolution under Eq.~\eqref{eqn.bob} then proceeds in a different direction. Each directional change is deterministic, but the kicks occur stochastically, and fluctuations in the time between kicks cause his Bloch vector to diffuse. The different components of Bob's dynamics are illustrated in Fig.~\ref{fig.braiding}(b)-(d).   

For generic braiding phases $0 \ll \theta \ll \pi$, we treat the resulting motion as a random walk. Thus, the average time between kicks is $T_\mathrm{kick}=1/P_2 \gamma_M$ and its variance is $ \Delta T_\mathrm{kick}^2=P_1/P_2^2 \gamma_M^2$. In the Zeno limit $P_2 \gamma_M\gg \Omega_\Delta$, the meridional (north-south) motion constantly reverses direction. Progress only occurs due to the variance of the kick intervals, i.e., the effective step size of the random walk is given by $\Omega \Delta T_\mathrm{kick}$. In the anti-Zeno limit, $P_2 \gamma_M\ll \Omega_\Delta$, the Bloch vector completes multiple revolutions with radius $\Omega/\Omega_\Delta$ between kicks, and the direction after each kick is nearly independent of the one before. The random step size is given by this radius and is independent of $T_\mathrm{Kick}$. Thus, the resulting escape rates are
\begin{align}
 \Gamma_\mathrm{anyon} \approx \begin{cases}
\frac{ P_1 \Omega^2}{P_2 \gamma_M}\quad & P_2 \gamma_M\gg \Omega_\Delta\quad \text{(Zeno),}\\
\frac{\Omega^2 }{\Omega_\Delta^2}P_2 \gamma_M & P_2 \gamma_M\ll \Omega_\Delta \quad \text{(anti-Zeno)}.
 \end{cases}
 \label{eqn.life}
\end{align} 
The divergence of the Zeno rate in the limit of full QPC transmission signals a breakdown of the diffusive treatment as the system reverts to ballistic motion. However, this limit cannot be realized experimentally, as a fully closed QPC does not permit anyons to tunnel between the inside and outside. 

\textit{Zeno effect in quantum Hall interferometers}.~
 To reveal Bob's dynamics, we interfere Alice's outgoing states by activating QPC2. Tunneling at this QPC is described by a scattering matrix $S_2$ analogous to Eq.~\eqref{eqn.smatrix} but without the $\hat N$ dependence, since Bob is assumed to tunnel only at QPC1. The reflection amplitudes of the two QPCs satisfy $r_2 r_1^* = e^{2 i \Phi_\mathrm{AB}} r_2^* r_1$, where the phase $\Phi_\mathrm{AB}$ includes kinetic terms, the Aharonov-Bohm phase, and statistical contributions from anyons that remain localized throughout.

In an OMZI, the final state is $|\tilde f \rangle=S_2 \hat S_1 |i\rangle $.
Alice's probability $P_i=|\langle \tilde f |D_i\rangle|^2$ of arriving at the drain $D_i$ contains a constant contribution and an oscillatory term 
\begin{align}
\delta P_{1,2}=\pm 2 t_1t_2r_1r_2( p_\mathrm{in} \cos[\Phi_\mathrm{AB} ] + p_\mathrm{out} \cos[\Phi_\mathrm{AB} -2 \theta]),
 \label{eqn.probability}
 \end{align}
where $p_\mathrm{in}$ and $p_\mathrm{out}$ are Bob's probabilities to be inside or outside the interferometer in the state $|B\rangle$. Their contributions add incoherently, as expected for two distinguishable states. Eq.~\eqref{eqn.probability} also applies for FPIs in the weak backscattering limit $|r_i| \ll 1$; for larger transmission rates, multiple loops result in a non-sinusoidal dependence on $\Phi_\mathrm{AB}$.

In addition to revealing Bob's dynamics, closing QPC2 also alters them. The matrix $S_2$ acts trivially on Bob but performs a rotation between Alice's states $|D_1\rangle$, $|D_2\rangle$, which is equivalent to changing the measurement basis. To illustrate this point, consider $50\%$ transmission at both QPCs with $r_i,t_i=\frac{1}{\sqrt{2}}$ and $\theta=\frac{\pi}{2}$. The final state in this case is\begin{align}
 |\tilde f\rangle = |D_1\rangle \otimes \hat N |B\rangle + |D_2\rangle \otimes (1-\hat N) |B\rangle,
\end{align}
which describes a projective measurement in which Alice's arrival at either drain conveys complete information about Bob's location. In this case, the interferometer realizes the canonical example of the quantum Zeno effect.

 \begin{figure}
 \centering
 \includegraphics[width=\linewidth]{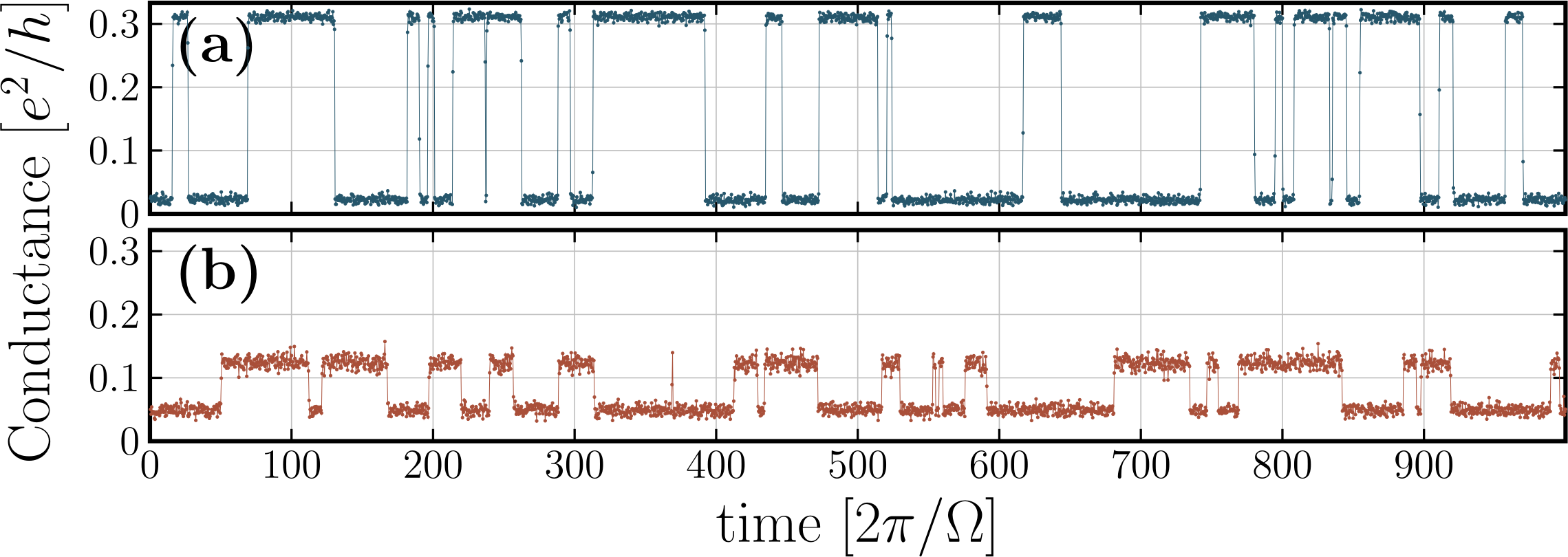}\\
 \includegraphics[height=3.4cm]{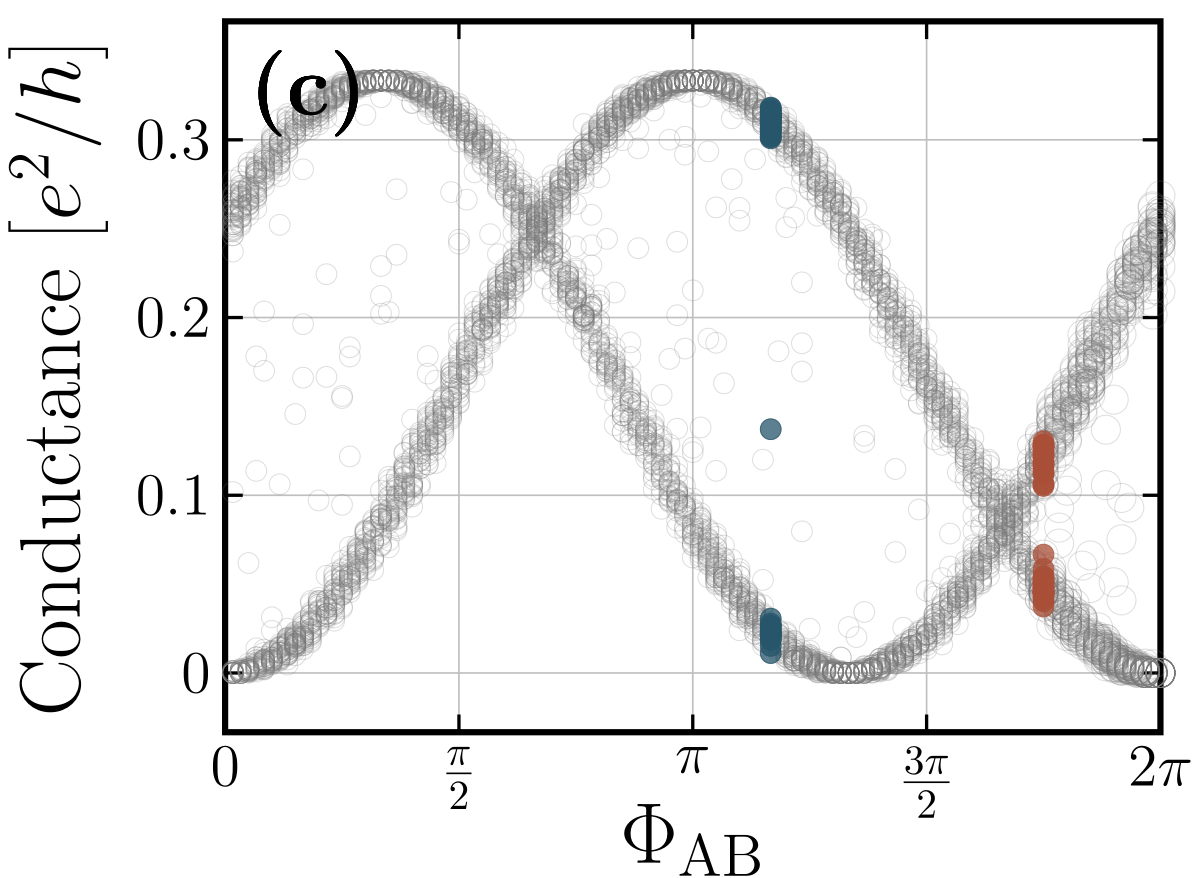}
 \includegraphics[height=3.4cm]{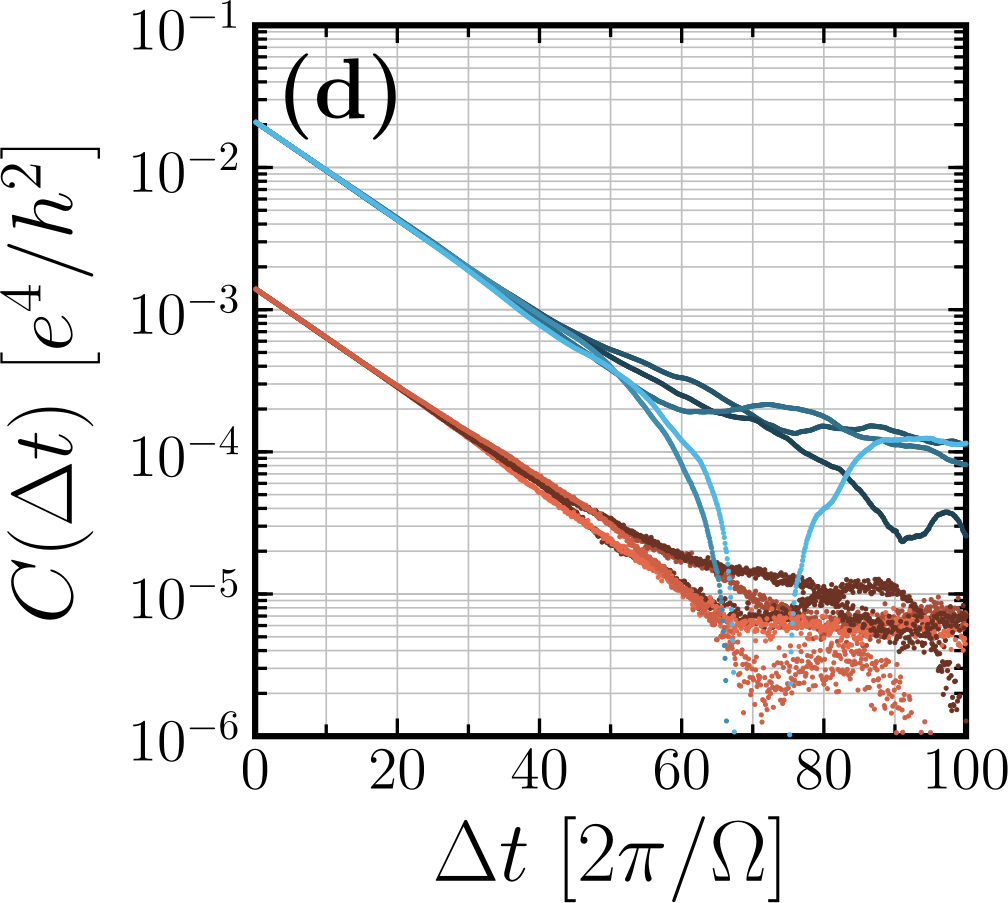}
 \caption{(a) and (b) Simulated time evolution of the conductance through an OMZI whose bulk realizes the $\nu=\frac{1}{3}$ Laughlin state with $50\%$ transmission at both QPCs. 
 The magnitudes of the jumps depend on the Aharonov-Bohm flux $\Phi_\mathrm{AB}$. (c) Conductance histogram as a function of $\Phi_\mathrm{AB}$. Orange and blue markers indicate the values chosen for the time traces. (d) The autocorrelation functions $C(\Delta t)$ for independent simulations with the same parameters as in (a) or (b). They exhibit an identical exponential decay before reaching the noise floor due to finite sampling time.}

 \label{fig.trace}
\end{figure}

\textit{Numerical results}.~
For generic parameter choices, Bob undergoes a combination of the unitary `kick' dynamics and partial projection. We simulate the time evolution of such systems in three alternating steps: (i) Bob undergoes unitary time evolution for a time $T_\mathrm{scatter}=\gamma_M^{-1}$, until disrupted by the next scattering event. (ii) The entangled final state $|\tilde f\rangle$ of Alice and Bob is constructed as described above Eq.~\eqref{eqn.probability}. (iii) Alice is measured projectively. The measurement outcome is selected randomly with the probabilities $P_{1,2}$, and Bob's state is updated accordingly. These steps are repeated $10^9$ times and the binary measurement outcomes $V(t)$ are recorded.

Each scattering event corresponds to a charge $e^*$ arriving at either drain. Conventional conductance measurements occur in the kHz-MHz range and average over time scales that are long compared to the arrival rate of individual carriers. Consequently, we perform a Gaussian filter with a chosen width of $100$ scattering events on $V(t)$ to relate it to the expected conductance. 

Results for the $\nu=\frac{1}{3}$ Laughlin state at a fixed current and with $\Omega_\Delta=0$ are presented in Fig.~\ref{fig.trace}. The average time between scattering events was $T_\mathrm{scatter}=10^{-3}\times\frac{2\pi}{\Omega}$, and both QPCs were set to $50\%$ transmission. Panels (a) and (b) show the time-dependent conductance for two different Aharonov-Bohm fluxes. At irregular intervals, the conductance jumps between two values determined by the cosines in Eq.~\eqref{eqn.probability} as shown in the histogram in panel (c). Bob's lifetime is extracted from the conductance autocorrelation function plotted in panel (d). The Aharonov-Bohm flux does not affect $\tau_\mathrm{anyon}$, even though the prefactors differ by an order of magnitude.

\begin{figure}
 \centering
 \includegraphics[width=\linewidth]{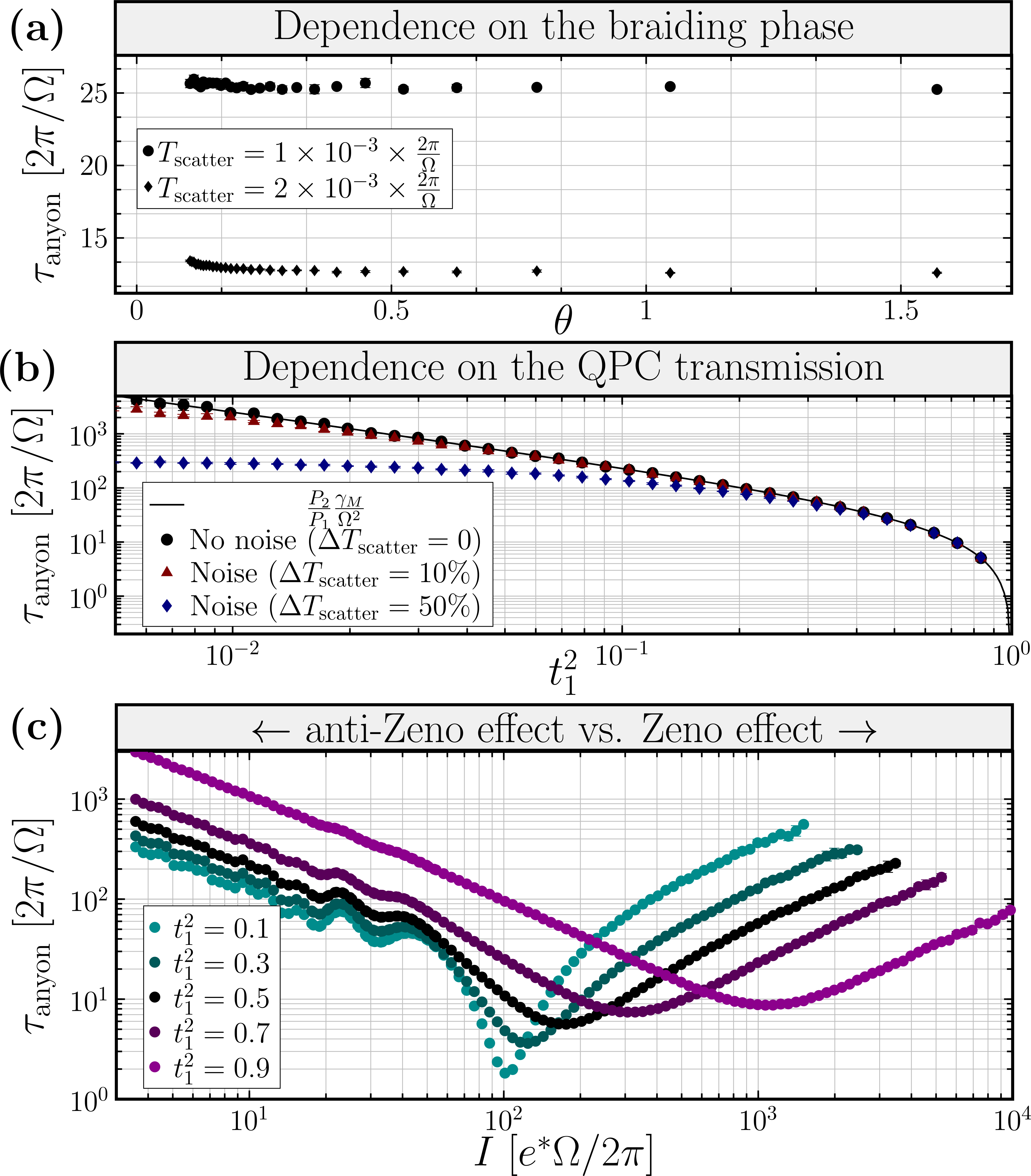} \caption{(a) The anyon lifetime depends only weakly on the braiding phase $\theta$. No data are shown for very small $\theta$, where $C(\Delta t)$ no longer follows an exponential decay. (b) The 
 dependence of Bob's lifetime on the QPC transmission follows Eq.~\ref{eqn.life}. Variations of 10\% in the measurement rate lead to slight deviations in the limit of a nearly closed QPC. (c) Crossover between the Zeno and anti-Zeno regimes as a function of the measurement rate. All data include error bars reflecting the small variations between independent simulations that differ in the measurement outcomes. }
 \label{fig.zeno}
\end{figure}

Figures~\ref{fig.zeno}(a) and (b) show how $\tau_\mathrm{anyon}$ in the Zeno regime depends on the braiding phase and the QPC transmission. As anticipated, there is no significant sensitivity to the braiding phase. Its variation with the QPC transmission is in good quantitative agreement with Eq.~\eqref{eqn.life}. In particular, $\tau_\mathrm{anyon}$ depends monotonously on the QPC transmission $t_1^2$, and not on the measurement strength $\sim (1-t_1^2)t_1^2$, which we consider a characteristic feature of the anyon Zeno effect. We also test how variations in the measurement intervals of $10\%$ and $50\%$, corresponding to a noisy current, affect the lifetime, and found that the differences are only significant at very low QPC transmissions.

Lastly, we turn to the crossover between Zeno and anti-Zeno regimes and perform simulations at different measurement rates. Figure~\ref{fig.zeno}(c) shows $\tau_\mathrm{anyon}$ as a function of $I$ for several QPC transmissions and $\Omega_\Delta = 50 \Omega$. The escape time \textit{increases} linearly with $I$ at large currents, while it \textit{decreases} at intermediate currents. The smooth crossover at large QPC transmissions is replaced by a sharp dip and oscillatory behavior at stronger reflection. This behavior reflects the unitary kick dynamics, which are strongest in this limit. In these simulations, we use a noisy current, which does not affect the Zeno limit but suppresses unphysical resonances in the anti-Zeno regime.

\textit{Discussion}.~Our main predictions of the anyon Zeno effect are: (i) The autocorrelation time \textit{increases} with the applied current at antidot settings where $|0\rangle$ and $|1\rangle$ are near degenerate (ii) When the energy difference between those states is large, the escape time $\tau_\text{anyon}$ is larger than in the first case, but \textit{decreases} with current. (iii) The autocorrelation time depends non-monotonically on the measurement strength and is largest for more reflective QPCs ($t_i^2 < r_i^2$). By contrast, the time-averaged probabilities for Bob to occupy either state are determined by thermodynamics and not directly affected by repeated measurements.

Our analysis assumed that Bob hops onto a well-defined state outside the interference loop. In practice, he may tunnel out on either side or directly into the gapless edge and propagate elsewhere to be replaced by a `new Bob' at a later time. Additionally, he may leave the interferometer to either side or even be replaced by Alice, as in the proposal of Ref.~\cite{kivelson2025}. However, we expect that the signatures (i)-(iii), which reflect generic properties of anyons, are robust. Moreover, the Zeno effect may also be operative in interferometers without an antidot. Provided Bob's bare tunneling rate (at $I=0$) is sufficiently small to be exceeded by $\gamma_M$, his lifetime is extended by the Zeno effect once a current is applied.

Time-dependent conductance measurements in interferometers without antidots have reported anyon dynamics on the second- or minute-scale \cite{werkmeister2024anyon, samuelson2024anyonic}, consistent with Zeno dynamics. Similar measurements in interferometers where bulk anyons can be controlled via a gate, together with an analysis of their current dependence, could provide direct evidence for the anyon Zeno effect. An observation of this phenomenon would represent a direct manifestation of anyonic braiding and introduce experimental control over the anyon dynamics. 

Finally, we briefly comment on non-Abelian quantum Hall states \cite{Moore_nonabelions_1991,Greiter_half_filled_1991,Wen_Non-Abelian_1991,Read_Beyond_1999,Stern_non_Abelian_2010}. For paired states at half-filling, the neutral-fermion parity within the interference loop can also affect the interference phase \cite{sarma2005,stern2006,bonderson2006,nayak_non-abelian_2008,rosenow2012}. Still, when either Alice or Bob is an Abelian $e/2$ particle, the parity is a mere spectator, and our analysis of the Abelian case applies unchanged. The Zeno dynamics in the case where both Alice and Bob are non-Abelian $e/4$ quasiparticles is directly sensitive to their non-Abelian statistics and will be an interesting subject for future research. 

We thank Yuval Ronen and Moty Heiblum for illuminating discussions that helped develop the idea, and the Israel Science Foundation (ISF) for funding this work under grant 3281/25.
\bibliography{ref}

@article{greenfield2025,
      title={A unified picture for quantum {zeno} and anti-{zeno} effects -- a review}, 
      author={Sacha Greenfield and Archana Kamal and Justin Dressel and Eli Levenson-Falk},
      year={2025},
      journal = {arXiv:2506.12679 [quant-ph]},
      eprint={2506.12679},
      archivePrefix={arXiv},
      primaryClass={quant-ph},
      url={https://arxiv.org/abs/2506.12679}, 
}

@article{KimAD,
  journal = {Philip {K}im, private communiation}
}

@article{YuvalAD,
 journal = {In preparation},
 author = {Kim, Jehyun
and Shaer, Amit
and Kumar, Ravi
and Ilin, Alexey
and Watanabe, Kenji
and Taniguchi, Takashi
and Mross, David
and Stern, Ady
and Ronen, Yuval}
}

@article{bonderson2006,
 title = {Detecting Non-{Abelian} Statistics in the $\ensuremath{\nu}=5/2$ Fractional Quantum {Hall} State},
 author = {Bonderson, Parsa and Kitaev, Alexei and Shtengel, Kirill},
 journal = {Phys. Rev. Lett.},
 volume = {96},
 issue = {1},
 pages = {016803},
 numpages = {4},
 year = {2006},
 month = {Jan},
 publisher = {American Physical Society},
 doi = {10.1103/PhysRevLett.96.016803},
 url = {https://link.aps.org/doi/10.1103/PhysRevLett.96.016803}
}

@article{deviatov,
 title = {Experimental realization of a {Fabry}-{Perot}-type interferometer by copropagating edge states in the quantum {Hall} regime},
 author = {Deviatov, E. V. and Lorke, A.},
 journal = {Phys. Rev. B},
 volume = {77},
 issue = {16},
 pages = {161302},
 numpages = {4},
 year = {2008},
 month = {Apr},
 publisher = {American Physical Society},
 doi = {10.1103/PhysRevB.77.161302},
 url = {https://link.aps.org/doi/10.1103/PhysRevB.77.161302}
}

@article{batra2023,
 title = {Anyonic {Mach}-{Zehnder} interferometer on a single edge of a two-dimensional electron gas},
 author = {Batra, Navketan and Wei, Zezhu and Vishveshwara, Smitha and Feldman, D. E.},
 journal = {Phys. Rev. B},
 volume = {108},
 issue = {24},
 pages = {L241302},
 numpages = {6},
 year = {2023},
 month = {Dec},
 publisher = {American Physical Society},
 doi = {10.1103/PhysRevB.108.L241302},
 url = {https://link.aps.org/doi/10.1103/PhysRevB.108.L241302}
}

@article{stern2006,
 title = {Proposed Experiments to Probe the Non-{Abelian} $\ensuremath{\nu}=5/2$ Quantum {Hall} State},
 author = {Stern, Ady and Halperin, Bertrand I.},
 journal = {Phys. Rev. Lett.},
 volume = {96},
 issue = {1},
 pages = {016802},
 numpages = {4},
 year = {2006},
 month = {Jan},
 publisher = {American Physical Society},
 doi = {10.1103/PhysRevLett.96.016802},
 url = {https://link.aps.org/doi/10.1103/PhysRevLett.96.016802}
}

@article{sarma2005,
 title = {Topologically Protected Qubits from a Possible Non-{Abelian} Fractional Quantum {Hall} State},
 author = {Das Sarma, Sankar and Freedman, Michael and Nayak, Chetan},
 journal = {Phys. Rev. Lett.},
 volume = {94},
 issue = {16},
 pages = {166802},
 numpages = {4},
 year = {2005},
 month = {Apr},
 publisher = {American Physical Society},
 doi = {10.1103/PhysRevLett.94.166802},
 url = {https://link.aps.org/doi/10.1103/PhysRevLett.94.166802}
}

@article{rosenow2012,
 title = {Telegraph noise and the {Fabry}-{Perot} quantum {Hall} interferometer},
 author = {Rosenow, B. and Simon, Steven H.},
 journal = {Phys. Rev. B},
 volume = {85},
 issue = {20},
 pages = {201302},
 numpages = {5},
 year = {2012},
 month = {May},
 publisher = {American Physical Society},
 doi = {10.1103/PhysRevB.85.201302},
 url = {https://link.aps.org/doi/10.1103/PhysRevB.85.201302}
}

@article{Levkivskyi2012,
 title = {Theory of fractional quantum {Hall} interferometers},
 author = {Levkivskyi, Ivan P. and Fr\"ohlich, J\"urg and Sukhorukov, Eugene V.},
 journal = {Phys. Rev. B},
 volume = {86},
 issue = {24},
 pages = {245105},
 numpages = {21},
 year = {2012},
 month = {Dec},
 publisher = {American Physical Society},
 doi = {10.1103/PhysRevB.86.245105},
 url = {https://link.aps.org/doi/10.1103/PhysRevB.86.245105}
}

@article{zhang2009,
 title = {Distinct signatures for {Coulomb} blockade and {Aharonov}-{{Bohm}} interference in electronic {Fabry}-{Perot} interferometers},
 author = {Zhang, Yiming and McClure, D. T. and Levenson-Falk, E. M. and Marcus, C. M. and Pfeiffer, L. N. and West, K. W.},
 journal = {Phys. Rev. B},
 volume = {79},
 issue = {24},
 pages = {241304},
 numpages = {4},
 year = {2009},
 month = {Jun},
 publisher = {American Physical Society},
 doi = {10.1103/PhysRevB.79.241304},
 url = {https://link.aps.org/doi/10.1103/PhysRevB.79.241304}
}

@article{ofek2010,
author = {Nissim Ofek and Aveek Bid and Moty Heiblum and Ady Stern and Vladimir Umansky and Diana Mahalu },
title = {Role of interactions in an electronic {Fabry}–{Perot} interferometer operating in the quantum {Hall} effect regime},
journal = {Proceedings of the National Academy of Sciences},
volume = {107},
number = {12},
pages = {5276-5281},
year = {2010},
doi = {10.1073/pnas.0912624107},
URL = {https://www.pnas.org/doi/abs/10.1073/pnas.0912624107},
eeprint = {https://www.pnas.org/doi/pdf/10.1073/pnas.0912624107}}

@article{halperin20211,
 title = {Theory of the {Fabry}-{P}\'erot quantum {Hall} interferometer},
 author = {Halperin, Bertrand I. and Stern, Ady and Neder, Izhar and Rosenow, Bernd},
 journal = {Phys. Rev. B},
 volume = {83},
 issue = {15},
 pages = {155440},
 numpages = {17},
 year = {2011},
 month = {Apr},
 publisher = {American Physical Society},
 doi = {10.1103/PhysRevB.83.155440},
 url = {https://link.aps.org/doi/10.1103/PhysRevB.83.155440}
}

@Article{Han2016,
author={Han, Cheolhee
and Park, Jinhong
and Gefen, Yuval
and Sim, H.-S.},
title={Topological vacuum bubbles by anyon braiding},
journal={Nature Communications},
year={2016},
month={Mar},
day={31},
volume={7},
number={1},
pages={11131},
issn={2041-1723},
doi={10.1038/ncomms11131},
url={https://doi.org/10.1038/ncomms11131}
}

@article{Wilczek1982,
 title = {Quantum Mechanics of Fractional-Spin Particles},
 author = {Wilczek, Frank},
 journal = {Phys. Rev. Lett.},
 volume = {49},
 issue = {14},
 pages = {957--959},
 numpages = {0},
 year = {1982},
 month = {Oct},
 publisher = {American Physical Society},
 doi = {10.1103/PhysRevLett.49.957},
 url = {https://link.aps.org/doi/10.1103/PhysRevLett.49.957}
}

@Article{Leinaas1977,
author={Leinaas, J. M.
and Myrheim, J.},
title={On the theory of identical particles},
journal={Il Nuovo Cimento B},
year={1977},
month={Jan},
day={01},
volume={37},
number={1},
pages={1-23},
issn={1826-9877},
doi={10.1007/BF02727953},
url={https://doi.org/10.1007/BF02727953}
}

@article{nakamura2020direct,
author={Nakamura, J.
and Liang, S.
and Gardner, G. C.
and Manfra, M. J.},
title={Direct observation of anyonic braiding statistics},
journal={Nature Physics},
year={2020},
month={Sep},
day={01},
volume={16},
number={9},
pages={931-936},
issn={1745-2481},
doi={10.1038/s41567-020-1019-1},
url={https://doi.org/10.1038/s41567-020-1019-1}
}

@article{giovannetti2008,
 title = {Multichannel architecture for electronic quantum {Hall} interferometry},
 author = {Giovannetti, Vittorio and Taddei, Fabio and Frustaglia, Diego and Fazio, Rosario},
 journal = {Phys. Rev. B},
 volume = {77},
 issue = {15},
 pages = {155320},
 numpages = {5},
 year = {2008},
 month = {Apr},
 publisher = {American Physical Society},
 doi = {10.1103/PhysRevB.77.155320},
 url = {https://link.aps.org/doi/10.1103/PhysRevB.77.155320}
}

@article{Chamon1997,
 title = {Two point-contact interferometer for quantum {Hall} systems},
 author = {de C. Chamon, C. and Freed, D. E. and Kivelson, S. A. and Sondhi, S. L. and Wen, X. G.},
 journal = {Phys. Rev. B},
 volume = {55},
 issue = {4},
 pages = {2331--2343},
 numpages = {0},
 year = {1997},
 month = {Jan},
 publisher = {American Physical Society},
 doi = {10.1103/PhysRevB.55.2331},
 url = {https://link.aps.org/doi/10.1103/PhysRevB.55.2331}
}

@article{facchi2001,
  title = {From the Quantum {zeno} to the Inverse Quantum {zeno} Effect},
  author = {Facchi, P. and Nakazato, H. and Pascazio, S.},
  journal = {Phys. Rev. Lett.},
  volume = {86},
  issue = {13},
  pages = {2699--2703},
  numpages = {0},
  year = {2001},
  month = {Mar},
  publisher = {American Physical Society},
  doi = {10.1103/PhysRevLett.86.2699},
  url = {https://link.aps.org/doi/10.1103/PhysRevLett.86.2699}
}

@article{kivelson2025,
  title = {Modified Interferometer to Measure Anyonic Braiding Statistics},
  author = {Kivelson, Steven A. and Murthy, Chaitanya},
  journal = {Phys. Rev. Lett.},
  volume = {135},
  issue = {12},
  pages = {126605},
  numpages = {6},
  year = {2025},
  month = {Sep},
  publisher = {American Physical Society},
  doi = {10.1103/x4w5-h3bb},
  url = {https://link.aps.org/doi/10.1103/x4w5-h3bb}
}

@Article{Buks1998,
author={Buks, E.
and Schuster, R.
and Heiblum, M.
and Mahalu, D.
and Umansky, V.},
title={Dephasing in electron interference by a `which-path' detector},
journal={Nature},
year={1998},
month={Feb},
day={01},
volume={391},
number={6670},
pages={871-874},
issn={1476-4687},
doi={10.1038/36057},
url={https://doi.org/10.1038/36057}
}

@article{aleiner1997,
  title = {Dephasing and the Orthogonality Catastrophe in Tunneling through a Quantum Dot: The ``Which Path?'' Interferometer},
  author = {Aleiner, I. L. and Wingreen, Ned S. and Meir, Yigal},
  journal = {Phys. Rev. Lett.},
  volume = {79},
  issue = {19},
  pages = {3740--3743},
  numpages = {0},
  year = {1997},
  month = {Nov},
  publisher = {American Physical Society},
  doi = {10.1103/PhysRevLett.79.3740},
  url = {https://link.aps.org/doi/10.1103/PhysRevLett.79.3740}
}

@article{stern2008,
title = {Anyons and the quantum {Hall} effect—A pedagogical review},
journal = {Annals of Physics},
volume = {323},
number = {1},
pages = {204-249},
year = {2008},
note = {January Special Issue 2008},
issn = {0003-4916},
doi = {https://doi.org/10.1016/j.aop.2007.10.008},
url = {https://www.sciencedirect.com/science/article/pii/S0003491607001674},
author = {Ady Stern}
}

@article{nakamura2023Fabry,
 title = {{Fabry}-P\'erot Interferometry at the $\ensuremath{\nu}=2/5$ Fractional Quantum {Hall} State},
 author = {Nakamura, J. and Liang, S. and Gardner, G. C. and Manfra, M. J.},
 journal = {Phys. Rev. X},
 volume = {13},
 issue = {4},
 pages = {041012},
 numpages = {11},
 year = {2023},
 month = {Oct},
 publisher = {American Physical Society},
 doi = {10.1103/PhysRevX.13.041012},
 url = {https://link.aps.org/doi/10.1103/PhysRevX.13.041012}
}

@article{kim2024aharonov,
author={Kim, Jehyun
and Dev, Himanshu
and Kumar, Ravi
and Ilin, Alexey
and Haug, Andr{\'e}
and Bhardwaj, Vishal
and Hong, Changki
and Watanabe, Kenji
and Taniguchi, Takashi
and Stern, Ady
and Ronen, Yuval},
title={{Aharonov}--{{Bohm}} interference and statistical phase-jump evolution in fractional quantum {Hall} states in bilayer graphene},
journal={Nature Nanotechnology},
year={2024},
month={Nov},
day={01},
volume={19},
number={11},
pages={1619-1626},
issn={1748-3395},
doi={10.1038/s41565-024-01751-w},
url={https://doi.org/10.1038/s41565-024-01751-w}
}

@article{samuelson2024anyonic,
 title={Anyonic statistics and slow quasiparticle dynamics in a graphene fractional quantum {{Hall}} interferometer},
 author={Samuelson, Noah L and Cohen, Liam A and Wang, Will and Blanch, Simon and Taniguchi, Takashi and Watanabe, Kenji and Zaletel, Michael P and Young, Andrea F},
 journal={cond-mat/2403.19628},
 year={2024}
}

@Article{kim2025aharonov,
author={Kim, Jehyun
and Dev, Himanshu
and Shaer, Amit
and Kumar, Ravi
and Ilin, Alexey
and Haug, Andr{\'e}
and Iskoz, Shelly
and Watanabe, Kenji
and Taniguchi, Takashi
and Mross, David F.
and Stern, Ady
and Ronen, Yuval},
title={{Aharonov}--{{Bohm}} interference in even-denominator fractional quantum {Hall} states},
journal={Nature},
year={2026},
month={Jan},
day={01},
volume={649},
number={8096},
pages={323-329},
issn={1476-4687},
doi={10.1038/s41586-025-09891-2},
url={https://doi.org/10.1038/s41586-025-09891-2}
}

@article{MisraSudarshan1977,
 author = {Misra, B. and Sudarshan, E. C. G.},
 title = {The {{zeno}}’s paradox in quantum theory},
 journal = {Journal of Mathematical Physics},
 volume = {18},
 number = {4},
 pages = {756-763},
 year = {1977},
 month = {04},
 issn = {0022-2488},
 doi = {10.1063/1.523304},
 url = {https://doi.org/10.1063/1.523304}
}

@article{Itano1990,
 author = {Itano, W. M. and Heinzen, D. J. and Bollinger, J. J. and Wineland, D. J.},
 title = {Quantum {{zeno}} effect},
 journal = {Physical Review A},
 volume = {41},
 number = {5},
 pages = {2295--2300},
 year = {1990},
 doi = {10.1103/PhysRevA.41.2295}
}

@article{FacchiPascazio2008,
doi = {10.1088/1751-8113/41/49/493001},
url = {https://doi.org/10.1088/1751-8113/41/49/493001},
year = {2008},
month = {oct},
publisher = {},
volume = {41},
number = {49},
pages = {493001},
author = {Facchi, P and Pascazio, S},
title = {Quantum {zeno} dynamics: mathematical and physical aspects},
journal = {Journal of Physics A: Mathematical and Theoretical}
}

@article{KofmanKurizki2000,
 author = {Kofman, A. G. and Kurizki, G.},
 title = {Acceleration of quantum decay processes by frequent observations},
 journal = {Nature},
 volume = {405},
 pages = {546--550},
 year = {2000},
 doi = {10.1038/35014537}
}

@article{SCHIEVE1989264,
title = {Numerical study of zeno and anti-zeno effects in a local potential model},
journal = {Physics Letters A},
volume = {136},
number = {6},
pages = {264-268},
year = {1989},
issn = {0375-9601},
doi = {https://doi.org/10.1016/0375-9601(89)90811-6},
url = {https://www.sciencedirect.com/science/article/pii/0375960189908116},
author = {W.C. Schieve and L.P. Horwitz and Jacob Levitan}
}

@article{LANE198491,
title = {Effect of frequent collisions on the decay of a state},
journal = {Physics Letters A},
volume = {104},
number = {2},
pages = {91-93},
year = {1984},
issn = {0375-9601},
doi = {https://doi.org/10.1016/0375-9601(84)90969-1},
url = {https://www.sciencedirect.com/science/article/pii/0375960184909691},
author = {A.M. Lane}
}

@article{nayak_non-Abelian_2008,
	title = {Non-{{Abelian}} anyons and topological quantum computation},
	volume = {80},
	issn = {0034-6861, 1539-0756},
	url = {https://link.aps.org/doi/10.1103/RevModPhys.80.1083},
		number = {3},
	urldate = {2017-09-16},
	journal = {Rev. Mod. Phys.},
	author = {Nayak, C. and Simon, S. H. and Stern, A. and Freedman, M. and Das Sarma, S.},
	month = sep,
	year = {2008},
	pages = {1083},
 doi = {10.1103/RevModPhys.80.1083}
}

@article{Moore_nonabelions_1991,
	title = {Nonabelions in the fractional quantum {{Hall}} effect},
	volume = {360},
	url = {http://www.sciencedirect.com/science/article/pii/055032139190407O},
	number = {2-3},
	urldate = {2017-09-22},
	journal = {Nucl. Phys. B},
	author = {Moore, G. and Read, N.},
	year = {1991},
	pages = {362},
}

@article{Greiter_half_filled_1991,
 title = {Paired {{Hall}} state at half filling},
 author = {Greiter, M. and Wen, X. G. and Wilczek, F.},
 journal = {Phys. Rev. Lett.},
 volume = {66},
 issue = {24},
 pages = {3205},
 numpages = {0},
 year = {1991},
 month = {Jun},
 publisher = {American Physical Society},
 doi = {10.1103/PhysRevLett.66.3205},
 url = {https://link.aps.org/doi/10.1103/PhysRevLett.66.3205}
}

@Article{Stern_non_Abelian_2010,
author={Stern, Ady},
title={Non-{Abelian} states of matter},
journal={Nature},
year={2010},
month={Mar},
day={01},
volume={464},
number={7286},
pages={187-193},
issn={1476-4687},
doi={10.1038/nature08915},
url={https://doi.org/10.1038/nature08915}
}

@article{Wen_Non-Abelian_1991,
 title = {Non-{{Abelian}} statistics in the fractional quantum {{Hall}} states},
 author = {Wen, X. G.},
 journal = {Phys. Rev. Lett.},
 volume = {66},
 issue = {6},
 pages = {802--805},
 numpages = {0},
 year = {1991},
 month = {Feb},
 publisher = {American Physical Society},
 doi = {10.1103/PhysRevLett.66.802},
 url = {https://link.aps.org/doi/10.1103/PhysRevLett.66.802}
}

@article{Read_Beyond_1999,
 title = {Beyond paired quantum {{Hall}} states: Parafermions and incompressible states in the first excited {Landau} level},
 author = {Read, N. and Rezayi, E.},
 journal = {Phys. Rev. B},
 volume = {59},
 issue = {12},
 pages = {8084--8092},
 numpages = {0},
 year = {1999},
 month = {Mar},
 publisher = {American Physical Society},
 doi = {10.1103/PhysRevB.59.8084},
 url = {https://link.aps.org/doi/10.1103/PhysRevB.59.8084}
}

@article{werkmeister2024anyon,
author = {T. Werkmeister and J. R. Ehrets and M. E. Wesson and D. H. Najafabadi and K. Watanabe and T. Taniguchi and B. I. Halperin and A. Yacoby and P. Kim },
title = {Anyon braiding and telegraph noise in a graphene interferometer},
journal = {Science},
volume = {388},
number = {6748},
pages = {730-735},
year = {2025},
doi = {10.1126/science.adp5015},
URL = {https://www.science.org/doi/abs/10.1126/science.adp5015},
eeeprint = {https://www.science.org/doi/pdf/10.1126/science.adp5015},
}

@Article{Ghosh_OMZI_2024,
author={Ghosh, B.
and Labendik, M.
and Musina, L.
and Umansky, V.
and Heiblum, M.
and Mross, D. F.},
title={Anyonic braiding in a chiral {Mach}--{Zehnder} interferometer},
journal={Nature Physics},
year={2025},
month={Jul},
day={02},
issn={1745-2481},
doi={10.1038/s41567-025-02960-3},
url={https://doi.org/10.1038/s41567-025-02960-3}
}

@Article{Ghosh_Coherent_Bunching_2024,
author={Ghosh, B.
and Labendik, M.
and Umansky, V.
and Heiblum, M.
and Mross, D. F.},
title={Coherent bunching of anyons and dissociation in an interference experiment},
journal={Nature},
year={2025},
month={Jun},
day={01},
volume={642},
number={8069},
pages={922-927},
issn={1476-4687},
doi={10.1038/s41586-025-09143-3},
url={https://doi.org/10.1038/s41586-025-09143-3}
}
\end{document}